\documentclass[12pt, letterpaper, twocolumn, aasmacros]{article}

\usepackage{pdfpages}

\usepackage[english]{babel}
\usepackage[utf8]{inputenc}

\usepackage{textcomp}

\usepackage[numbers]{natbib}
\usepackage[letterpaper, margin=1in]{geometry}

\usepackage{fancyhdr}
\usepackage{palatino}
\usepackage{titlesec}
\titleformat{\section}{\normalfont\fontsize{12}{15}\bfseries}{\thesection}{1em}{}
\titleformat{\subsection}{\normalfont\fontsize{12}{15}\itshape}{\thesubsection}{1em}{}
\titlespacing\section{0pt}{12pt plus 4pt minus 2pt}{0pt plus 2pt minus 2pt}
\titlespacing\subsection{0pt}{12pt plus 4pt minus 2pt}{0pt plus 2pt minus 2pt}

\usepackage{hyperref}

\usepackage{graphicx}
\usepackage{dcolumn}
\usepackage{bm}

\usepackage{wrapfig}
\usepackage[most]{tcolorbox}

\def\ao{Appl.~Opt.}           
\def\aap{A\&A}                

\NewTColorBox{MissionBox}{ s O{!htbp} }{%
floatplacement={#2},
IfBooleanTF={#1}{float*,width=\textwidth}{float},
colback=yellow!5!white,
colframe=yellow!50!black, 
colbacktitle=yellow!75!black,
title=\textbf{Box 1 - LISA Mission Overview},
after=\par\nointerlineskip,
lower separated=false
}

\NewTColorBox{SciBox}{ s O{!htbp} }{%
  floatplacement={#2},
  IfBooleanTF={#1}{float*,width=\textwidth}{float},
  colback=yellow!5!white,
  colframe=yellow!50!black, 
  colbacktitle=yellow!75!black,
  title=\textbf{Box 2 - LISA Science Objectives },
  lower separated=false
  }
  
\NewTColorBox{AccelBox}{ s O{!htbp} }{%
floatplacement={#2},
IfBooleanTF={#1}{float*,width=\textwidth}{float},
colback=yellow!5!white,colframe=yellow!50!black,
colbacktitle=yellow!75!black,
title=\textbf{Box 3 - Near-perfect free-fall in LISA Pathfinder},
lower separated=false
}

\NewTColorBox{TechBox}{ s O{!htbp} }{%
floatplacement={#2},
IfBooleanTF={#1}{float*,width=\textwidth}{float},
colback=yellow!5!white,
colframe=yellow!50!black, 
colbacktitle=yellow!75!black,
title=\textbf{Box 4 - NASA-supported Technology Development for LISA},
lower separated=false
}

\begin{document}

\null
\includepdf{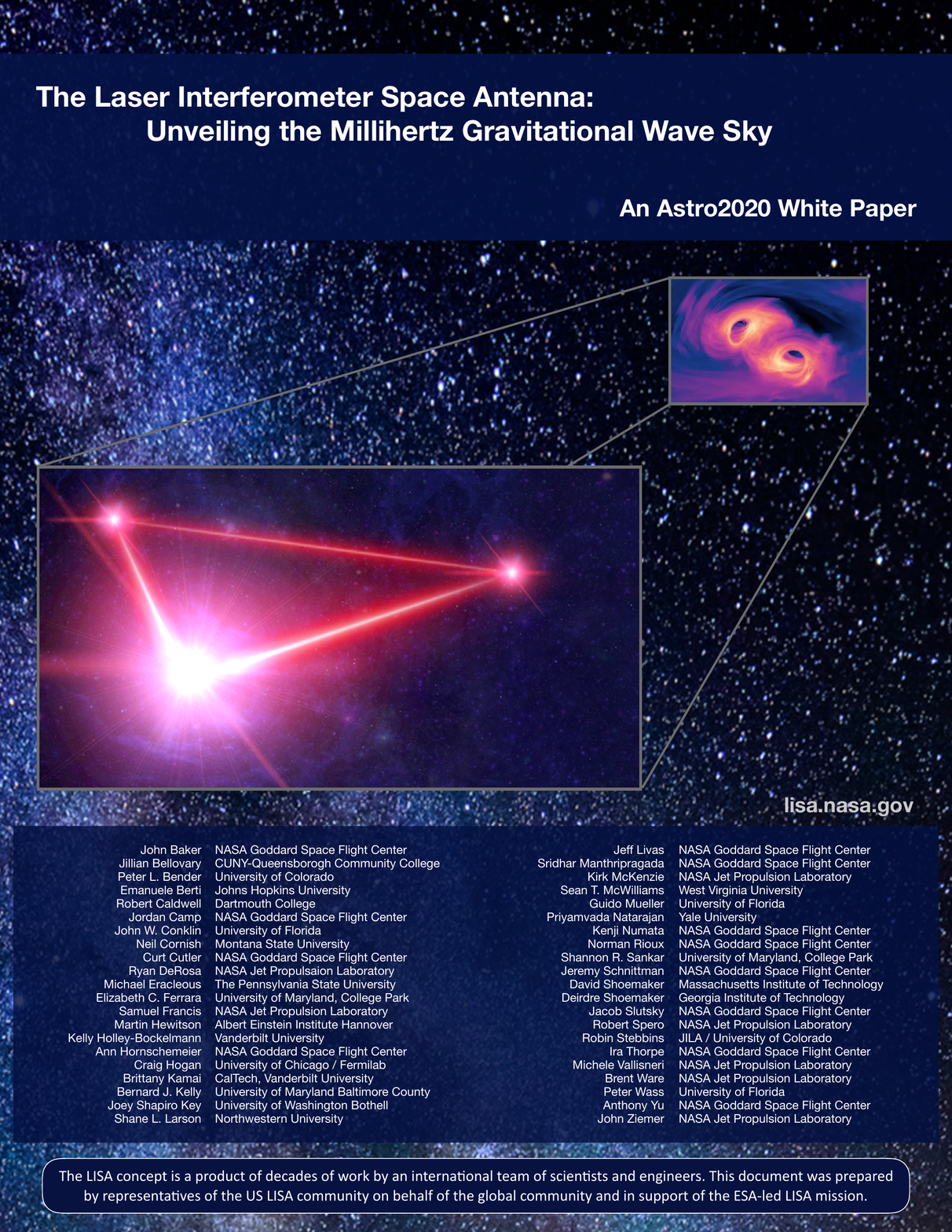}
\pagenumbering{gobble}
\setcounter{page}{1}
\pagenumbering{arabic}

\section{\label{sec:exec}Executive Summary}
The first terrestrial Gravitational Wave (GW) interferometers~\cite{GW150914,2018arXiv181112907T} have dramatically underscored the scientific value of observing the Universe through an entirely different window -- and of folding this new channel of information with traditional astronomical data for a multimessenger view~\cite{2017ApJ...848L..12A,2018PhRvL.121p1101A,2017Natur.551...80K,2017Natur.551...85A}. \textbf{The Laser Interferometer Space Antenna (LISA) will broaden the reach of GW astronomy by conducting the first survey of the millihertz GW sky}, detecting tens of thousands of individual astrophysical sources ranging from white-dwarf binaries in our own galaxy to mergers of massive black holes (MBHs) at redshifts extending beyond the epoch of reionization. \textbf{These observations will inform -- and transform -- our understanding of the end state of stellar evolution, MBH birth, and the co-evolution of galaxies and black holes through cosmic time.}  LISA also has the potential to detect GW emission from elusive astrophysical sources such as intermediate-mass black holes as well as exotic cosmological sources such as inflationary fields and cosmic string cusps\footnote{LISA Astro2020 Science whitepapers available at: \url{lisa.nasa.gov/documentsCWP.html}}. 

\begin{MissionBox}[h!]
\noindent
\textbf{Science Objective:} All-sky survey of millihertz gravitational waves
\medskip
\newline
\textbf{Measurement Concept:} Long-baseline optical interferometry between drag-free test masses
\medskip
\newline
\textbf{Orbit:} Heliocentric 2.5 Mkm triangular constellation, 20$^\circ$ Earth-trailing 
\medskip
\newline
\textbf{Launch:} Early/mid 2030s, Ariane 6.4
\vspace{-5pt}
\newline
\textbf{Lifetime:} 1.5 yr transfer, 1 yr commissioning, 4 yrs science, $\leq$ 6 yrs extension
\medskip
\newline
\textbf{Cost:}  Total mission: Large ($>$\$1.5B); US share: Medium (\$500M - \$1.5B)
\medskip
\newline
\textbf{Partners:} European Space Agency (lead), ESA Member States, NASA 

\end{MissionBox}

LISA is now in Phase A as a European Space Agency (ESA) led mission with significant contributions anticipated from several ESA member states and NASA. \textbf{The mission concept retains all essential features of the NASA/ESA LISA mission that was ranked as the 3rd priority for Large-class missions in the 2010 Decadal Survey}~\cite{NWNH}, including the full three-arm triangular configuration that measures GW polarization and improves robustness. Since that ranking, \textbf{LISA's technical readiness has been greatly advanced through two flight demonstrations}: the ESA-led LISA Pathfinder mission (2015-2017) and the Laser Ranging Instrument on board the US/German Gravity Recovery And Climate Explorer Follow-On mission (2018-). 
\newline
The Midterm Assessment of the 2010 Decadal Survey recommended that the US participate as a ``strong technical and scientific partner" in an ESA-led LISA mission~\cite{Midterm}. NASA is currently supporting  pre-project activities to support a range of potential contributions to LISA including instruments, spacecraft elements, and science analysis. The currently envisioned scale of these contributions is at the lower end of the medium-scale cost range identified by Astro2020 (\$500M - \$1.5B). \textbf{A recommendation for an upscope in US participation in LISA would provide opportunities to more fully exploit heritage from prior US investments, balance technical and programmatic risks across the  partnership, and expand opportunities for future US leadership in this new field of astronomy.}

\section{\label{sec:science}Key Science Goals and Objectives}
The scientific case for GW observations in the millihertz band is well-summarized in \emph{The Gravitational Universe}~\cite{gravUniverse}, the document which formed the basis for ESA's selection of GW astronomy as the science theme for the 3rd Large-class mission of the Cosmic Vision Programme in 2013.  While some elements of the LISA science case overlap with those for GW observations in other bands~\cite{2019BAAS...51c.232S}, notably higher-frequency observations with terrestrial interferometers and lower-frequency observations with Pulsar Timing Arrays, a space-based facility such as LISA is uniquely capable of answering a number of pressing and fundamental questions in astrophysics.

The LISA Science Objectives are formally documented in the Science Requirements Document (SciRD)~\cite{SciRD}. The SciRD identifies eight LISA Science Objectives (SOs, see Box 2) -- broad questions in astrophysics which can be addressed through GW astronomy. The SOs are used to derive mission and instrument requirements including requirements on the sensitivity of LISA to GW strain as well as other factors including total observing time and data latency (for certain classes of EM counterpart investigations).  

\begin{SciBox}[h!]
\noindent
\textbf{SO 1:} Study the formation \& evolution of compact binary stars in the Milky Way
\medskip
\newline
\textbf{SO 2:} Trace the origin, growth \& merger history of MBHs 
\medskip
\newline
\textbf{SO 3:} Probe the dynamics of dense nuclear clusters using EMRIs
\medskip
\newline
\textbf{SO 4:} Understand the astrophysics of stellar origin black holes
\medskip
\newline
\textbf{SO 5:} Explore the fundamental nature of gravity \& black holes
\medskip
\newline
\textbf{SO 6:} Probe the rate of expansion of the Universe
\medskip
\newline
\textbf{SO 7:} Understand stochastic GW backgrounds \& their implications for the early Universe and TeV-scale particle physics
\medskip
\newline
\textbf{SO 8:} Search for GW bursts and unforeseen sources
\end{SciBox}

\begin{figure}[ht]
\includegraphics[width=\columnwidth]{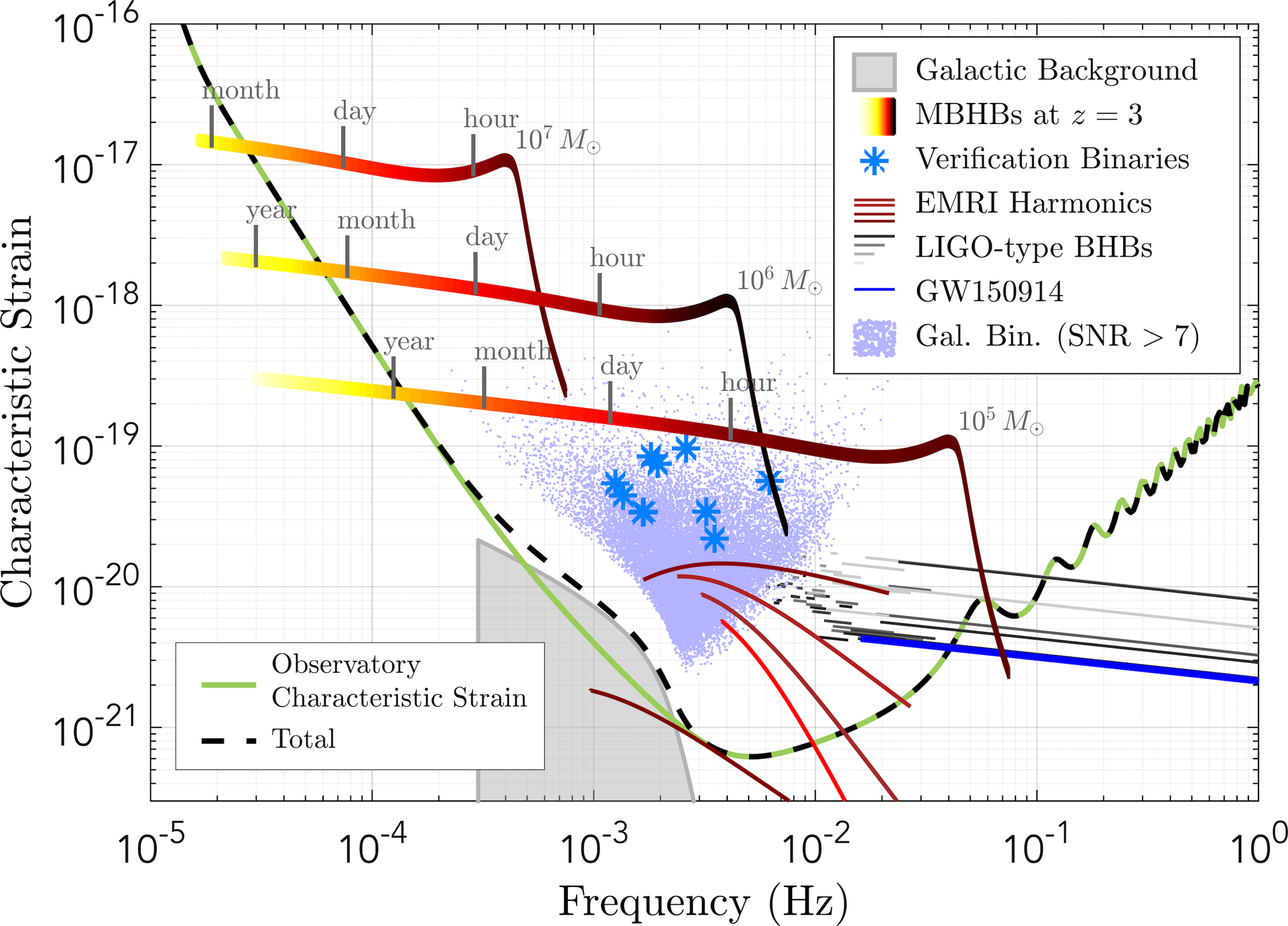}
\caption{Representative examples of LISA sources compared with the instrument sensitivity. Sources and instrument sensitivity are plotted as frequency spectra of characteristic GW strain. All sources are observed simultaneously and individually extracted through a global fit of the LISA time-series data. \emph{Figure 1 from ~\cite{L3proposal}}. \label{fig:sources} }
\end{figure}

The broad range of scientific targets is made possible by the abundance and diversity of astrophysical sources expected for LISA. Figure \ref{fig:sources} shows a comparison of several representative LISA sources with the sensitivity limit of the instrument. Data are plotted as spectral amplitudes of GW strain -- a dimensionless number characterizing the amplitude of the spacetime stretching caused by GWs passing through the detector. The `characteristic strain' is used to account for variations in observation time between transient and persistent signals~\cite{2015CQGra..32a5014M}. Sensitivity to astrophysical sources  is primarily limited by instrument noise (green trace), which will be discussed in more detail in Section \ref{sec:mission}. The representative sources in Figure \ref{fig:sources} are, in order of typical distance from nearest to most distant.

\medskip
\noindent
\emph{Compact Binaries in the Milky Way}
\newline
LISA will be sensitive to millions of binary systems of compact objects (white dwarfs, neutron stars, or black holes). Tens of thousands will be individually resolved with measured masses, orbital parameters, and 3D locations~\cite{2017MNRAS.470.1894K, 2019arXiv190700014L} (blue dots in Figure \ref{fig:sources}).  The remaining systems contribute to an unresolved foreground (grey shaded region) which limits sensitivity to other GW sources (black dashed line). Several ``verification binaries'' are already known through EM observations~\cite{2018MNRAS.480..302K}, providing guaranteed multimessenger sources (large blue asterisks). LISA observations will provide insight on stellar populations and evolution as well as the dynamics of compact binaries ~\cite{2019arXiv190305583L}.

\medskip
\noindent
\emph{Black Hole Binaries}
\newline
Black hole binaries with component masses in the approximate range of 10-100 $M_\odot$, such as have been observed with terrestrial GW observatories~\cite{2018arXiv181112907T}, will also be observable by LISA at earlier epochs in their evolution. Unlike lighter binary systems, these systems evolve appreciably during LISA's observing lifetime (grey and black traces, blue trace is GW150914), occasionally exiting the LISA band before rapidly evolving to merger in the $\sim100\,$Hz band, raising the intriguing possibility of \emph{multiband} GW observations~\cite{2016PhRvL.116w1102S}. Applications include tests of gravity and fundamental physics through cross-comparison of GW measurements in the millihertz and audio bands ~\cite{2019BAAS...51c.109C}.
\newline

\medskip
\noindent
\emph{Extreme Mass-Ratio Inspirals (EMRIs)}
\newline
A unique class of signal in the millihertz band is the capture of black holes or neutron stars by MBHs in the local Universe ($z\leq 2$). The large difference in mass between the two objects results in a highly complex orbit with multiple frequency components simultaneously evolving (five reddish traces representing a single EMRI at z=1.2). LISA will observe tens to hundreds of these EMRI events,  yielding one of the most precise possible tests of General Relativity in the strong-field regime and also providing unique insight into the demographics and dynamics of the high mass end of the population of objects in nuclear clusters of galaxies like the Milky Way~\cite{2019BAAS...51c..42B}. LISA may also be able to detect GWs from the capture, and eventual disruption, of individual WDs by MBHs in the nearby Universe, leading to an exciting new multimessenger source~\cite{2019BAAS...51c..10E}.

\medskip
\noindent
\emph{Massive black hole mergers}
\newline
LISA will detect hundreds of black hole mergers with signal-to-noise ratios of 10$-$10$^4$ and redshift of 1$-$30. Multi-colored traces show three example equal-mass MBH mergers at z=3 which sweep across the LISA band from low to high frequencies with time before merger, as indicated on the track.  LISA will provide opportunities to probe the birth and growth of massive black holes and their host galaxies at redshift ranges and for halo mass ranges that are not readily accessible with other techniques~\cite{2019BAAS...51c..73N,2019arXiv190306867C}. Sky localization to $\mathcal{O}(10\,\textrm{arcmin})$ by merger for the highest-SNR (and most nearby) systems, supporting multimessenger observations to provide insight into the astrophysical environments of merging MBHs as well as independent measurements of cosmological expansion via standard sirens~\cite{2019BAAS...51c.123B}.
\newline

\medskip
\noindent
\emph{Exotic Sources}
\newline
The potential for discovery may be the strongest motivation for making observations in this as-yet-unobserved window on the Universe ~\cite{2019BAAS...51c..76C}. Possibilities include both astrophysical sources (e.g. intermediate-mass black holes~\cite{2019BAAS...51c.175B}) and  cosmological sources (e.g. GW backgrounds from inflation or early-universe phase
transitions, and cosmic string bursts, etc.~\cite{2019BAAS...51c..67C}).

\section{\label{sec:mission}Mission Overview}
Observing in the millihertz band requires a space-based facility, much as observing in parts of the infrared electromagnetic spectrum requires going to space. Terrestrial GW detectors are limited to higher frequencies by gravitational coupling to seismic density fluctuations that are increasingly severe at low frequencies. More fundamental is the physical size of the detectors themselves, which are not sufficiently sensitive to the long wavelengths of  millihertz GWs. In contrast, LISA can be placed in an orbit far from Earth where the thermal, magnetic, and gravitational environment is far more stable and the observatory can be expanded to the million-km baselines that maximize sensitivity to the GW signals of interest.

\begin{figure}[ht]
\includegraphics[width=\columnwidth]{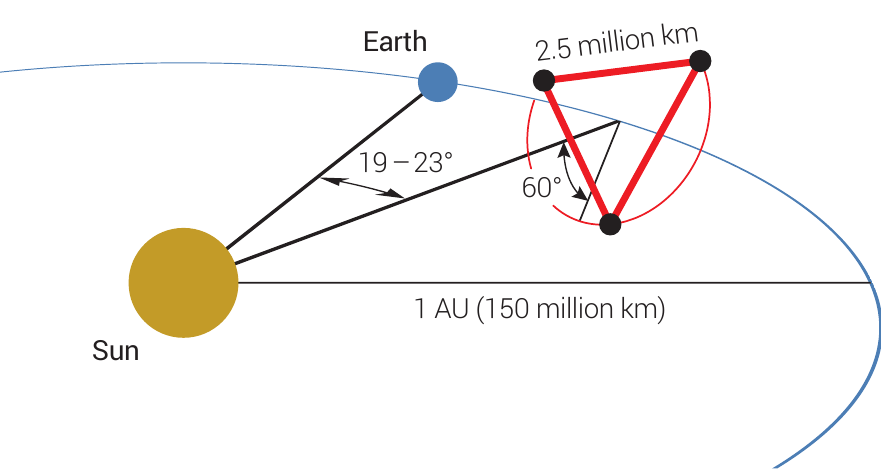}
\caption{Orbital configuration of the LISA mission. The 2.5Mkm triangular constellation is inclined to the ecliptic by 60$^\circ$ and undergoes a cartwheeling motion once per orbit. \label{fig:orbits} }
\end{figure}

\begin{AccelBox}*[h!]
\begin{wrapfigure}{r}{0.6\textwidth}
\vspace{-20pt}
\includegraphics[width=0.6\textwidth]{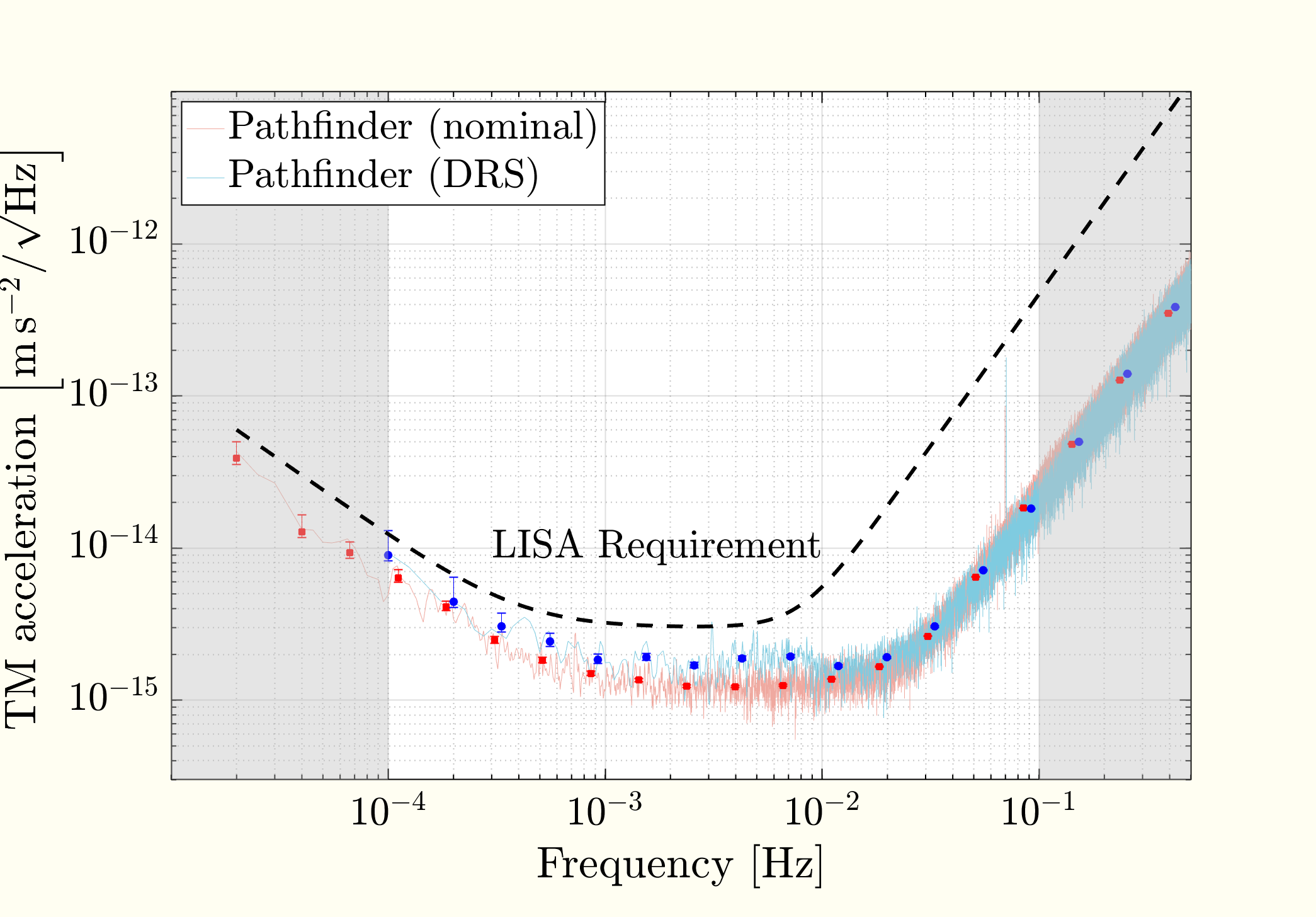}
\vspace{-35pt}
\end{wrapfigure}
LISA Pathfinder placed a single 2kg Au-Pt test mass in a LISA-like drag-free configuration and employed a second identical test mass and an optical interferometer as a low-noise witness sensor. The figure shows the amplitude spectral density of the equivalent single test mass residual acceleration noise for Pathfinder in the nominal configuration~\cite{2018PhRvL.120f1101A} (red curve). Also shown is  an analogous measurement made over a shorter time and with correspondingly decreased statistics in a configuration where the NASA-supplied colloidal micronewton thrusters were used in place of the cold-gas thrusters ~\cite{2018PhRvD..98j2005A} (light blue curve). Both of these spectra outperform the LISA MRD requirement (black dashed line). In addition to characterizing acceleration noise performance, Pathfinder provided experience in sub-picometer interferometric metrology of free-flying test masses, non-contact charge control~\cite{2018PhRvD..98f2001A}, precision drag-free control using micropropulsion~\cite{2019PhRvD..99h2001A,lpfColdGas, 2018PhRvD..98j2005A}, in-flight thermal diagnostics in the LISA band~\cite{2019MNRAS.486.3368A}, and other techniques and technologies relevant to LISA. 
\end{AccelBox} 

\subsection{\label{sec:missionDesign}LISA mission design}
At the most basic level, the LISA measurement concept parallels that which has been successfully employed on LIGO and other terrestrial GW interferometers~\cite{2015CQGra..32g4001L,2015CQGra..32b4001A}. A set of test masses are arranged across widely-separated baselines and the proper distance between these masses is monitored using optical interferometry for fluctuations caused by passing GWs. In the case of LISA, the test masses are arranged in three pairs, with each pair hosted in a spacecraft placed at one vertex of an approximately-equilateral triangle with a side-length of 2.5 Mkm. The resulting constellation forms a triangle that is inclined at 60$^o$ with respect to the ecliptic plane and undergoes a cartwheeling motion with one rotation in the constellation plane per orbit about the Sun. This configuration, which is depicted in Figure \ref{fig:orbits}, remains passively stable over the lifetime of the LISA mission (1yr commissioning \& calibration + 4 yrs science operations + 6 yrs potential extended operations)~\cite{1997CQGra..14.1405F}. 

\subsection{\label{sec:drivers}Performance Drivers}
The sensitivity of the observatory is chiefly determined by two performance metrics: the level of imperfection of the test-mass free-fall and the precision with which changes in the test mass separation can be measured.  The requirements on each of these are defined in the LISA Mission Requirements Document (MRD). Flowing these top-level mission requirements down to requirements on individual subsystems and components is a major focus of the current mission formulation activities. This process begins with a detailed performance model that accounts for each physical effect that contributes to either test mass acceleration or displacement metrology noise. These models can then be used to evaluate different instrument architectures and develop an error budget at each level of the system. The key performance drivers for test mass acceleration noise are residual gas pressure in the test mass cavity, control of electrostatic charges on the test mass, stability of the electrostatic suspension used to control the test mass in the non-measurement degrees of freedom, and careful control of the magnetic, thermal, and gravitational environment of the spacecraft. The displacement measurement is fundamentally limited by photon shot noise, which is in turn determined by laser power, telescope diameter, and arm length. Reaching this fundamental limit requires mitigation of technical noises, the largest of which is laser frequency noise which couples into the measurement through the unequal arms of the LISA constellation. Laser noise is mitigated through a combination of active stabilization of the laser frequency and application of a post-processing technique known as Time-Delay Interferometry~\cite{1999ApJ...527..814A, 2004PhRvD..70h1101S, 2005PhRvD..72d2003V}. Beyond laser frequency noise, the key performance drivers for the displacement performance are thermomechanical stability of the optical structures, mitigation of scattered light, and geometrical errors that lead to coupling of spacecraft jitter into the displacement measurement. 

LISA's technical readiness has taken significant steps forward since the 2010 Decadal Survey. Much of this progress is due to two in-flight demonstrations that validated key aspects of LISA's measurement concept and several critical technologies. LISA Pathfinder (2015-2017) was an ESA-led mission with the express purpose of increasing technical readiness for LISA.  The instrument included a pair of representative LISA test masses, one of which was placed in a LISA-like configuration of drag-free flight and the other of which was used as a low-noise witness for measuring residual accelerations of the primary test mass (see Box 3). The Gravity Recovery And Climate Explorer Follow-On (2018-) is the replacement for the highly successful US/German GRACE mission.  Most relevant to LISA, GRACE-FO includes the Laser Ranging Instrument (LRI), a laser interferometer  which measures the inter-satellite distance in parallel with the primary microwave ranging instrument. Recent results from the LRI have demonstrated nanometer-level interferometric ranging over a 210 km link~\cite{2019arXiv190700104A}, meeting the design goals of LRI which are relaxed from LISA due to the larger size of GRACE-FO's geodesy signal relative to LISA's GW signals. Nevertheless, LRI provides flight heritage for key LISA components such as photoreceivers, phase measurement systems, and laser control systems as well as valuable experience with operational activities such as link acquisition.

\subsection{\label{sec:data}Science Analysis}
The combined effects of LISA's all-sky sensitivity and the long observational duration of typical LISA sources results in $\mathcal{O}(10^4)$ individual signals with signal-to-noise  ratios (SNRs) above the detection threshold existing simultaneously in the LISA data stream. Extracting each of these sources accurately and efficiently is a critical part of the LISA measurement effort.  The primary tools for addressing this challenge are matched-filtering and Markov-Chain Monte Carlo techniques, which have been successfully employed by terrestrial GW interferometers to extract and characterize signals. LISA adds the complexity of overlapping signals but benefits from the fact that each of the signals accumulates many cycles in GW phase during the measurement, helping to resolve overlapping signals and more precisely measure astrophysical parameters~\cite{2009CQGra..26i4024V}. Addressing the LISA data analysis challenge requires a coordinated effort of a team with a diverse set of expertise including source astrophysics, gravitational waveforms, deep knowledge of the instrument, and expertise in matched-filtering searches. The international LISA community has worked to bring individual efforts together in a set of data challenge activities, in which simulated LISA data are generated with a known set of sources using a common set of simulation tools, and various groups work to analyze that data before gathering together to compare and discuss results.  The first series of such exercises, known as the Mock LISA Data Challenges (MLDCs), were carried out by the joint NASA-ESA science team in the late 2000s~\cite{2006AIPC..873..619A,2008CQGra..25k4037B}. LISA Data challenge exercises have recently been renewed\footnote{\url{lisa-ldc.lal.in2p3.fr/ldc}}, incorporating lessons learned from the MLDCs as well as advances in the understanding of both the astrophysical signals and the LISA instrument, that allow for more faithful simulated data sets.

\section{\label{sec:org}Organization and Current Status}

\begin{TechBox}*[ht]
NASA is currently supporting the development of five separate technologies for potential contribution to LISA. The strategy for selecting these technologies, which is laid out in detail the Interim Report of the L3 Study Team~\cite{L3ST}, is to balance impact and insight into the LISA system, heritage from prior NASA investments, and tractability of interfaces. The NASA Study Office Technology Plan~\cite{NASAtech} describes in detail the development strategy for each of these technologies to reach ISO TRL 6 by Mission Adoption (2023). The table below briefly introduces them:
\begin{center}
\begin{tabular}[t]{ p{1.2in} | p{1.6in} | p{2.2in}}
\textbf{Technology} & \textbf{Role in LISA} & \textbf{Development Strategy} \\
\hline
Telescope & 
Efficiently deliver optical power across long baselines & 
Control dimensional stability using low-expansion materials and stable thermal design.  \\
\hline
Laser &
Provide light for primary interferometric measurement &
MOPA architecture utilizing NPRO technology from LISA Pathfinder and LRI \\
\hline
Charge Management &
Control electric charge on the test masses using UV light &
Build on Pathfinder heritage; replace Hg lamps with UV LEDs as light source \\
\hline
Micropropulsion & 
Precision attitude/position control of the spacecraft &
Leverage ST7 heritage, increase reliability and lifetime \\
\hline
Phase Measurement Systems &
Acquire primary science, auxiliary, and control-loop error signals. &
Build off LRI experience; add LISA-specific functionality \\
\end{tabular}
\end{center}
\end{TechBox}

ESA's Science Programme Committee selected \emph{The Gravitational Universe} (millihertz GW astronomy) as the science theme for the 3rd large-class mission in the Cosmic Vision Programme (L3) in 2013. Following the early successes of LISA Pathfinder~\cite{2016PhRvL.116w1101A} and the historic first observations of GWs by LIGO~\cite{GW150914}, a call for mission concepts was issued by ESA in 2016. A European-US team of scientists responded with a proposal for LISA in early 2017~\cite{L3proposal}, which was subsequently selected by ESA in June 2017. As is common for ESA missions, contributions to both flight hardware and science support are anticipated from ESA Member States as well as international partners. The LISA Consortium\footnote{\url{www.lisamission.org}} was subsequently formed to support the development of the payload and to coordinate the efforts of the international research community in areas of data analysis and science exploitation. 

\begin{figure*}[h!]
\includegraphics[width=\textwidth]{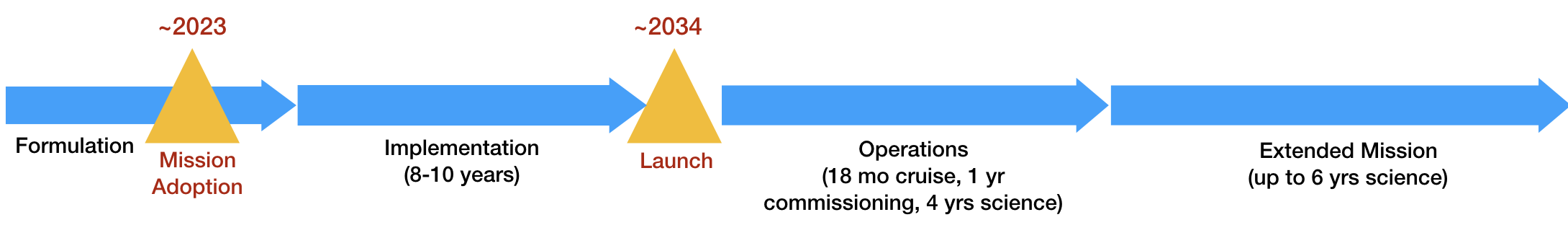}
\caption{An overview of the current LISA schedule including major project milestones. \label{fig:schedule} }
\end{figure*}

In 2015, NASA convened a team of US experts in LISA science and technology known as the L3 Study Team to study potential opportunities for US contribution to an ESA-led LISA mission~\cite{L3ST}. Following the strong recommendations for increased US participation in LISA by the 2016 Midterm Assessment of New Worlds, New Horizons~\cite{Midterm}, NASA established the NASA LISA Study Office (NLSO) to coordinate technical interchange with the ESA Study Team and its partners, consolidate technology development activities supporting potential US contributions, and engage with the scientific community in the US and Europe.  The L3ST was replaced by the NASA LISA Study Team (NLST) after the 2017 mission selection.

\medskip
The NLSO is located within the Physics of the Cosmos Program Office of the NASA Astrophysics Division and is led by the Goddard Space Flight Center with substantial contributions from the Jet Propulsion Laboratory, the Marshall Space Flight Center, and the University of Florida. The near-term goal of the NLSO is to identify a set of potential US contributions to LISA and assess the merit, risk, and cost of each contribution. Part of this effort includes developing a set of technologies to a level that makes them viable, low-risk contributions at the time that a final arrangement is made (Box 4). The Study Office will transition to a Phase A project once a final set of roles and responsibilities has been negotiated between ESA, NASA, and the partner ESA Member States. This agreement will be made at or before ESA's adoption milestone (see additional discussion in Section \ref{sec:schedule}) and will be influenced by technical readiness, suitability of interfaces, and available budgets among all partners.

A critical aspect of LISA is that it is effectively a single instrument distributed across a constellation of three spacecraft.  While it is common in space missions to have different organizations providing different instruments, it is not common to have these separately-sourced units interacting with one another in as intricate a way as LISA demands.  All of the LISA partner organizations recognize this fact and are closely cooperating at both technical and programmatic levels. Under a Letter of Agreement signed between NASA and ESA in 2019, the NASA and ESA Study Offices, and by extension ESA's European partners, are able to share technical information that allows the LISA design to proceed in a coordinated fashion. NLSO personnel have been invited to participate in a number of ESA and LISA Consortium-led activities, including progress meetings with industrial contractors performing  detailed design studies of the spacecraft and payload. A Systems Engineering Office (SEO) has been established to facilitate technical interchange among experts from across the partnership, and to harmonize processes for the development of requirements and interfaces across programmatic boundaries.   

\section{\label{sec:schedule}Schedule}

A simplified LISA schedule is shown in Figure \ref{fig:schedule}.  LISA is currently in the early stages of formulation as a Phase A Study at ESA and a pre-Phase A Study within NASA. ESA is currently conducting a mid Phase A review intended to close major architectural trades and identify any critical needs in technology development that require attention. This will be followed by a Mission Formulation Review as the gate review to Phase B. The next major milestone for LISA is Mission Adoption, a critical decision point in the ESA framework which authorizes the mission to proceed through the final stages of formulation and the implementation phase. In this sense it is analogous to the Confirmation milestone in the NASA framework. Mission Adoption is currently targeted for 2023. After Mission Adoption, industrial contracts for the implementation phase will be negotiated and the final formulation and implementation activities can begin. These phases typically last 8-10 years for an ESA L-class mission, leading to a launch date in the early 2030s, consistent with, or perhaps slightly in advance of, the target 2034 launch date for L3 in the ESA Cosmic Vision planning document.
As a junior partner, the NASA project schedule will track the ESA schedule where possible. The most significant discrepancies are in the formulation phase, when NASA will remain in pre-Phase A until the final set of NASA roles and responsibilities are negotiated with ESA and other European partner agencies. Input from the Decadal Survey, assuming it adheres to its late-2020/early-2021 estimate for release, will have an opportunity to influence this negotiation. Most importantly, all NASA technology development activities are being managed to a schedule that is consistent with ESA's guidelines, namely achieving an ISO technology readiness level (TRL) of 5/6 on all critical items prior to Mission Adoption. Further details on NASA's development schedule can be found in the Technology Development Plan~\cite{NASAtech}.

\section{\label{sec:cost}Cost}
As a junior partner contributing to a mission led by an international partner, the cost considerations for LISA are somewhat different than for other missions under consideration by Astro2020. While total-project cost estimates were previously made by both the NASA LISA project of the 2000s as well as by independent cost estimators~\cite{NAP12006,NWNH}, those cost estimates are of limited value for assessing the costs of an ESA-led mission, which operates under different financial conditions. Based on the ESA L-class cost cap of \texteuro 1.05 B and the expected level of contributions from European National agencies and NASA, it is reasonable to assume that the total LISA mission cost will lie in the ``large'' category identified in the Astro2020 APC guidelines ($> \$1.5\,\textrm{B}$). The size of the NASA contribution under any conceivable partnership arrangement would be in the ``medium'' category ($\$500\,\textrm{M}-\$1.5\,\textrm{B}$). This cost would include both the value of the hardware deliverables to Europe as well as contributions to the science ground segment, US Guest Investigator programs, and NASA project overhead including management, systems engineering, project science, and mission assurance.  The NLSO has developed and continues to refine cost estimates for the project lifecycle cost (LCC) of each potential hardware contribution as well as the science participation and project overhead activities. These LCC estimates are used to validate potential contribution scenarios against budget constraints.  A memo describing the LCC estimates and their underlying assumptions can be provided to the Astro2020 committee upon request.

\section{\label{sec:upscopes}Scenarios for US Participation}
LISA is a single scientific instrument that is distributed across a constellation of three spacecraft and which conducts an all-sky survey resulting in a single data set containing a mixture of all sources. Furthermore, the LISA science and mission requirements have been established and formulation activities are proceeding. The motivations for an upscope to the US LISA contribution are not to make the instrument more sensitive but rather to contribute to overall mission success and to boost US participation in a compelling new field in astronomy. The currently envisaged US contribution to LISA is at the lower end of the ``medium" scale identified by Astro2020 (\$500M-\$1.5B), which will enable the US to provide a subset of the instrument components described in Box 4 as well as contribute to the LISA science analysis efforts.  An upscope would expand the range of potential hardware contributions, allowing more complete utilization of the significant US-based investments in LISA and related efforts. Additionally, it would enable increased participation by the US community in LISA data analysis and science exploitation.  The specific set of US contributions will depend on a number of factors including technical readiness, compatibility with European partners, and available budgets. Here we present three broad categories of contributions which could be enabled by an upscope.

\medskip
\noindent
\emph{Engineering of Instrument Subsystems}
\newline 
Careful and deliberate systems engineering is key to the success of precision measurement apparatus such as LISA. While LISA is a distributed instrument, there are arrangements of instrument subsystems that minimize complexity of interfaces. Taking responsibility for one of these subsystems would be the most effective way for the US to mitigate risk through hardware contributions.  The most logical approach would be to build such a subsystem around one of the component-level technologies under development by NASA (see Box 4). In such a scenario, the US could additionally provide component-level contributions to the additional, European-led instrument subsystems.

\medskip
\noindent
\emph{US Science Facilities}
\newline
Analysis and scientific exploitation of mission data is typically funded by ESA Member States for ESA missions, and the same arrangement is being planned for LISA. An important part of the NASA LISA Project will be a Guest Investigator / Science Center facility which will fulfill a similar role for the US-based research community. In addition to implementing the US role in the project-level data analysis (e.g. similar to the arrangement on the ESA-led Planck and Herschel missions), such a facility would provide outside users with access to mission data at a variety of levels, as well as tools to facilitate working with LISA data and for combining LISA data with other facilities in multimessenger investigations. The scope of such a facility, as well as Guest Investigator grants to carry out research using LISA data, will be directly influenced by the scale of the NASA effort. 

\medskip
\noindent
\emph{Propulsion Alternatives}
\newline
In addition to GW strain sensitivity, the most determinative factor in LISA's science performance is mission lifetime. The three major determinants of mission lifetime are stability of the orbits, reliability of the spacecraft and instrument systems, and amount of propellant in the micropropulsion system.  The NASA-developed colloidal micronewton thrusters (CMNTs), performed successfully on LISA Pathfinder~\cite{2018PhRvD..98j2005A} and may enable a significantly lower system mass. Provided a mission architecture that effectively incorporated CMNTs could be developed, such a contribution could significantly enhance LISA science return by reducing system mass and increasing the propellant-limited lifetime. 
\newline
\medskip
\newline
 In many ways, LISA represents a unique opportunity for NASA as the junior partner -- \textbf{full participation in the science of a flagship-scale mission with a medium-scale investment}. A robust US contribution to LISA will further the success of this groundbreaking mission and will provide engagement and leadership opportunities for current and future members of the US science community.  

\clearpage
\onecolumn{\bibliography{LISAAstro2020}}

\end{document}